%% file: main.tex
\documentclass{article}
\usepackage{spconf,amsmath,graphicx}

\usepackage{xcolor}
\usepackage{multirow}
\usepackage{float}
\usepackage{cite}
\usepackage{amssymb}
\usepackage{algorithm}  
\usepackage{algpseudocode}  
\usepackage{booktabs,multirow}
\usepackage{adjustbox}
\usepackage{yhmath}
\usepackage{url}

\title{\textit{\textbf{PHONEix}}: Acoustic Feature Processing Strategy for \\ Enhanced Singing Pronunciation with Phoneme Distribution Predictor}

\name{
    Yuning Wu$^{1}$, 
    Jiatong Shi$^{2}$, 
    Tao Qian$^{1}$, 
    Dongji Gao$^{3}$, 
    Qin Jin$^{1*}$\thanks{*Corresponding author.}
}
\address{
    $^{1}$ School of Information, Renmin University of China \\
	$^{2}$ Language Technologies Institute, Carnegie Mellon University \\
	$^{3}$ Center for Language and Speech Processing, Johns Hopkins University \\
    \small{\texttt{\{yuningwu, qiantao, qjin\}@ruc.edu.cn}, \texttt{jiatongs@cs.cmu.edu}}
}

\begin{document}
\ninept
\maketitle

\begin{abstract}
Singing voice synthesis (SVS), as a specific task for generating the vocal singing voice from a music score, has drawn much attention in recent years. SVS faces the challenge that the singing has various pronunciation flexibility conditioned on the same music score. Most of the previous works of SVS can not well handle the misalignment between the music score and actual singing. In this paper, we propose an acoustic feature processing strategy, named \textbf{\textit{PHONEix}}, with a phoneme distribution predictor, to alleviate the gap between the music score and the singing voice, which can be easily adopted in different SVS systems. Extensive experiments in various settings demonstrate the effectiveness of our \textbf{\textit{PHONEix}} in both objective and subjective evaluations. 
\end{abstract}


\input{sec-intro}

\input{sec-method}

\input{sec-exp}

\section{Conclusion and Future Work}

In this work, we introduce for the first time an improved AFP strategy \textit{\textbf{PHONEix}} to an SVS system. It helps narrow the gap between theoretical music score input and actual singing. In it, a novel \textbf{phoneme distribution predictor} is integrated to give phoneme proportions in each note. Therefore, the phoneme duration becomes a learnable feature to fit the complex pronunciation in singing. The prior encoder module can be quickly adapted to different acoustic models. If given a large SVS dataset, it would be promising work to pre-train a phoneme distribution predictor to annotate phoneme time sequences in raw datasets.

\section{Acknowledgement}
This work was supported by the National Natural Science Foundation of China (No. 62072462) and the National Key R\&D Program of China (No. 2020AAA0108600).

\vfill\pagebreak
\ninept
\bibliographystyle{IEEEbib}
\bibliography{refs}

\end{document}

%% file: sec-intro.tex
\section{Introduction}
\label{sec:intro}

SVS aims to generate a singing voice from a music score that contains pitch and duration information organized by notes with corresponding lyrics~\cite{cook1996singing}. Different from the similar task of text-to-speech (TTS), it is difficult to obtain large public singing datasets for SVS \cite{baseline1, guo2022singaug, ren2020deepsinger}, due to the copyright restrictions and the strict recording environment requirements. Moreover, singing has higher variability than spoken language, as singers have the flexibility to make changes to scores, making singing more natural and pleasing. For example, the pronunciation of the same word can vary significantly due to pitch and tempo changes in singing. 
Therefore, it is challenging to learn the pronunciation of lyrics and make the singing match the melodic changes in the music score, especially on small-scale SVS corpora. Recently, a work \cite{Karaoker} trains a singing-data-free SVS model on speech datasets to imitate the voice of a singing template. Nevertheless, a common practice in current score-based SVS systems is to adopt acoustic feature processing (AFP) strategies in the acoustic model to enhance the learning of singing pronunciation based on limited SVS data.

\begin{figure}[t]
	\centering
	\includegraphics[width=1.0\columnwidth]{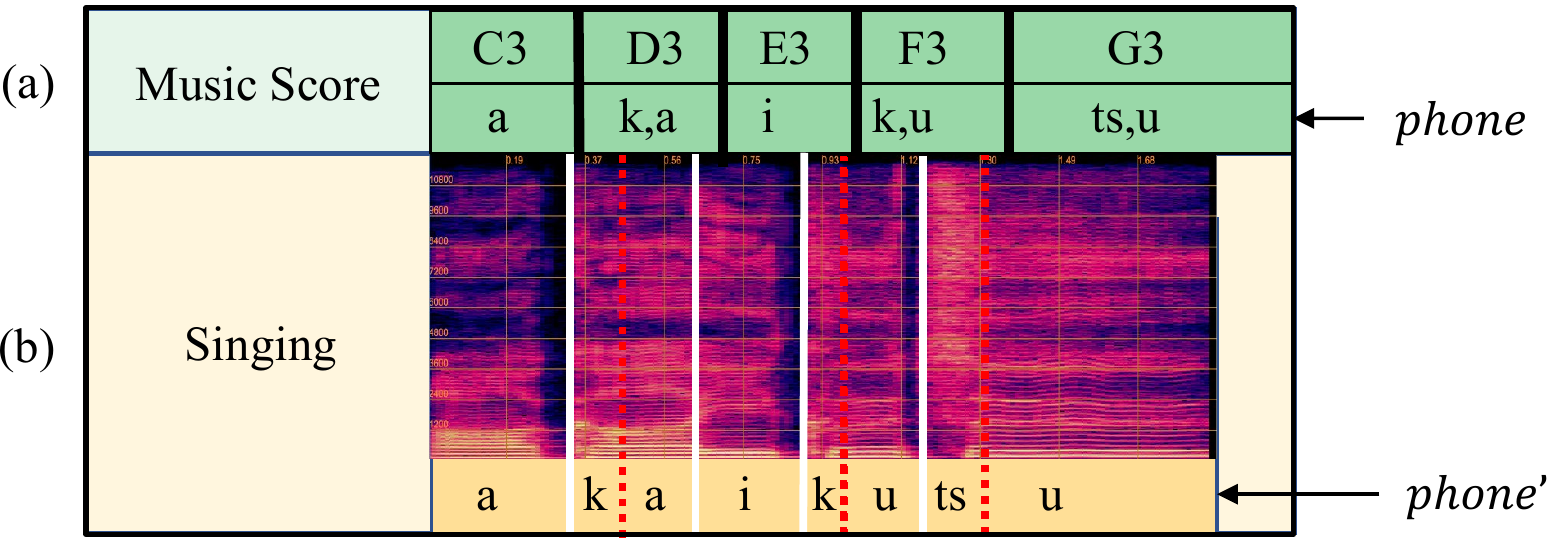}
	\caption{\small (a) Music score input. Black bold vertical lines denote note splits. (b) Manually labeled phoneme time sequence. White bold lines indicate syllable splits and red dotted lines indicate phoneme splits in each syllable.}
	\label{fig:input}
\end{figure}

\begin{figure*}[t]
	\centering
	\includegraphics[width=1.0\linewidth]{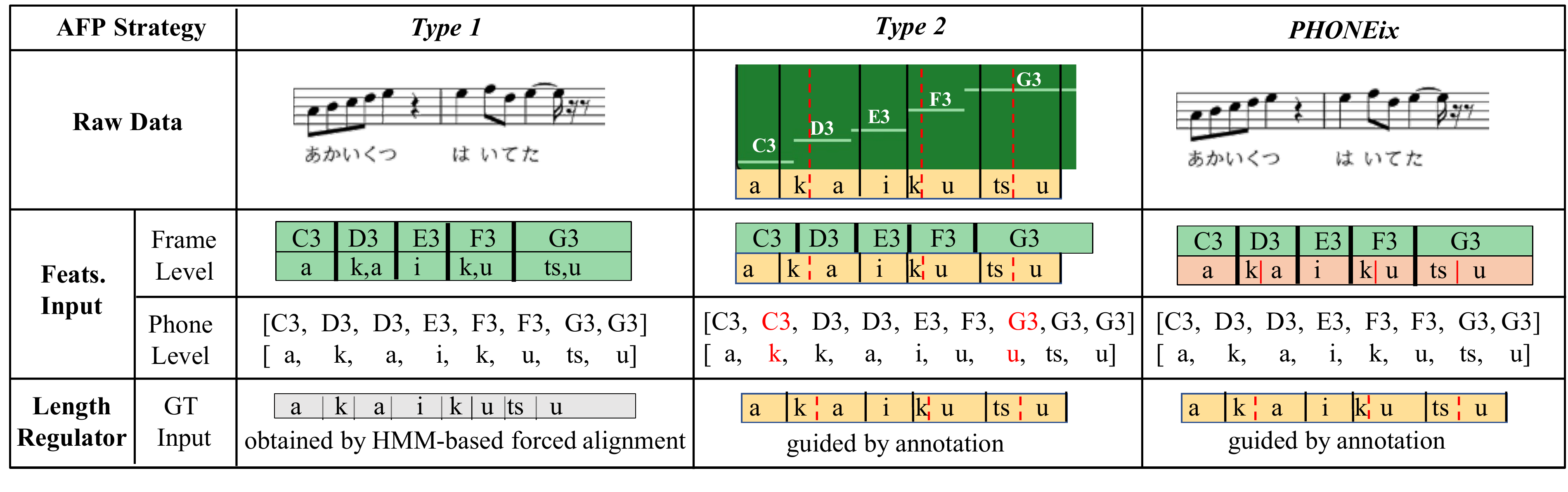}
  \vspace{-15pt}
	\caption{\small Different AFP strategy. Details of \textit{\textbf{Type 1}} and \textit{\textbf{Type 2}} are introduced in Sec.~\ref{sec:intro} and \textit{\textbf{PHONEix}} are in ~\ref{ssec: Acoustic Feature Processing Strategy}. Blocks of different colors represent phoneme duration. The green blocks are from the music score and the yellow blocks are from annotation. The gray block is obtained by HMM-based forced alignment. The orange block is predicted by the phoneme distribution predictor. In column \textit{\textbf{Type 2}} strategy and the red lines are division of vowels and consonants. The red phonemes in phone-level feature input are redundant phonemes and missing phonemes caused by misalignments between notes and annotation.}
	\label{fig:AFP}

 \end{figure*}
 
\begin{figure}[t]
	\centering
	\includegraphics[width=1.0\columnwidth]{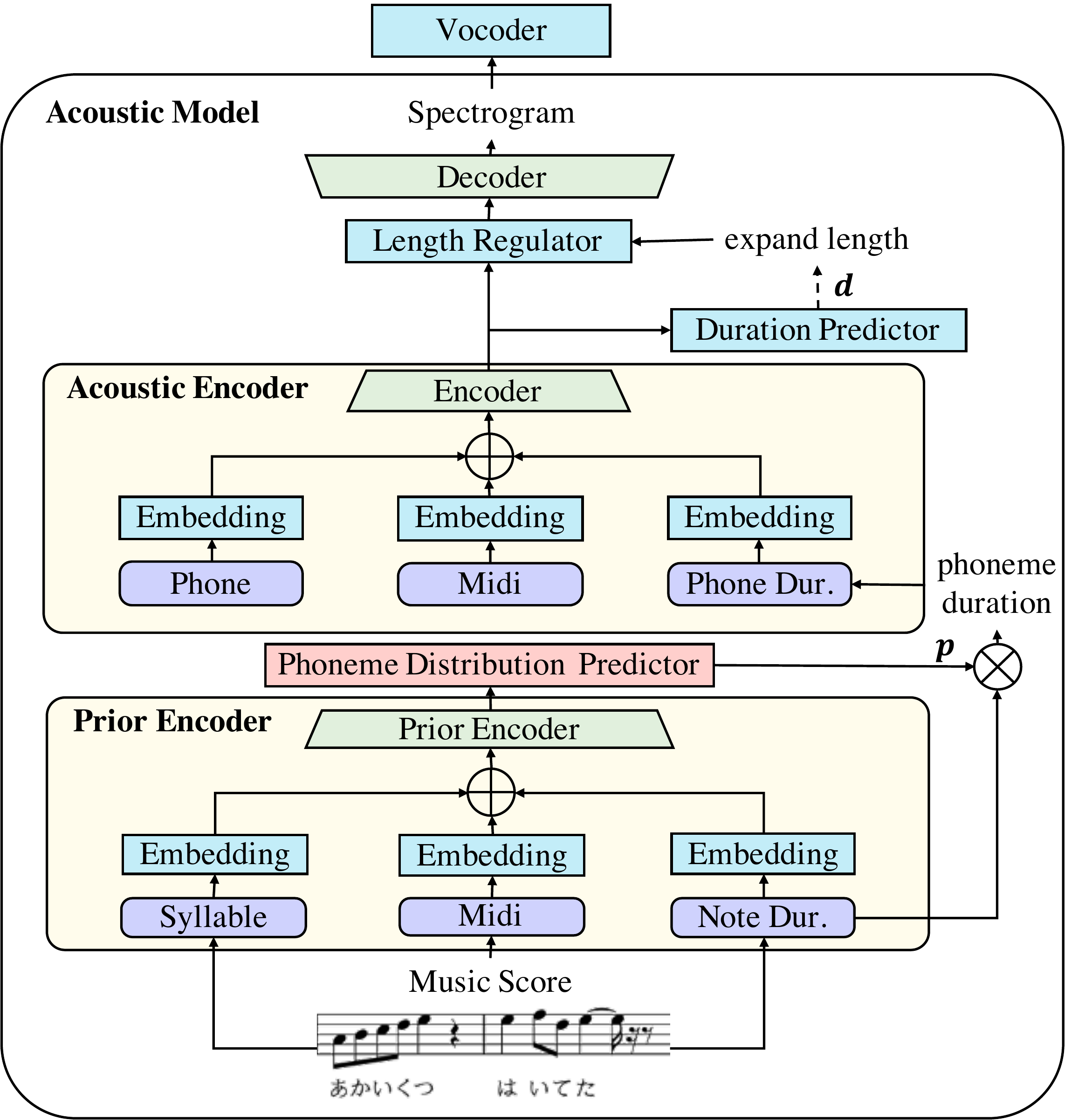}
	\caption{\small The SVS system architecture with \textit{\textbf{PHONEix}} AFP strategy. Details are introduced in Sec.~\ref{ssec: overall framework}. The green blocks can be adjusted to accommodate different acoustic models.}
	\label{fig:model}
\end{figure}




Current AFP strategies for SVS acoustic models can be roughly categorized into two types. 1) \textbf{\textit{Type~1}} AFP strategy is to use the original note pitch and whole note duration from the music score (i.e., Fig.~\ref{fig:input}~(a)) as the 
pitch and duration input, with no distinction between vowels and consonants. 
To further distinguish vowels and consonants, a duration predictor is built to produce fine-grained phoneme-level duration, which is trained based on supervision calculated by force-alignment \cite{oura2010recent, chen2020hifisinger, liu2022diffsinger, hono2021sinsy, yi2019singing, wang2022mr}, heuristics \cite{tae2021mlp, choi2022melody, lu2020xiaoicesing, baseline2} etc. 
The advantage of this type of feature processing strategy is that the input phoneme and pitch sequence are strictly aligned at the note level based on the music score.
However, as illustrated in Fig.~\ref{fig:input}, there is a gap between the \textit{\textbf{Type~1}} music score and the actual singing voice in phoneme duration distribution. Acoustic models are not able to adapt the distribution with only music scores. 
2) \textbf{\textit{Type~2}} AFP strategy is to use a labeled phoneme duration sequence (i.e. Fig.~\ref{fig:input}~(b)) and the note sequence from the music score (i.e. Fig.~\ref{fig:input}~(a)) as input \cite{zhang2022susing, lee2021n, zhuang2021litesing, liu2021vibrato, zhang2022wesinger, wang2022singing, gu2021bytesing, VISinger, kim2018korean,nakamura2019singing,hono2018recent}. The labeled phoneme duration sequence provides a supervision signal for duration predictor to predict the actual duration of phonemes and notes given \textit{\textbf{Type~1}} phoneme and note duration from the music score. 
The disadvantage is that the labeled phoneme duration sequence is annotated based on the singing voice and it is not exactly aligned with the music score. 
This misalignment will lead to redundant phonemes and missing phonemes, which will hurt the generation from the actual singing voice. Moreover, different from the training phase, the actual phoneme duration is unavailable during inference, such discrepancy between training and inference adversely leads to a great restriction of application scenarios for SVS. Overall, the divergence between music score and actual singing cannot be well solved in the acoustic model by these two types of strategies. 

To mitigate the gap between the music score and actual singing, 
we propose a new AFP strategy, named \textit{\textbf{PHONEix}}, for the SVS acoustic model. Specifically, the acoustic model accepts the music score as input, and then the proposed \textbf{phoneme distribution predictor} learns the phoneme durations, adapting to the actual pronunciation. Finally, the aligned \textit{\textbf{Type 1}} score features (phoneme, pitch, phoneme duration) are encoded and then length regulated under the guidance of the actual phoneme duration to address the gap (illustrated in Fig.~\ref{fig:AFP}). In summary, the contributions of this work include:
\begin{itemize}
    \item We propose a new AFP strategy \textit{\textbf{PHONEix}} for SVS, which narrows the gap between the music score and the singing voice by feeding \textit{\textbf{Type 1}} duration and actual duration successively into the acoustic model. \textit{\textbf{PHONEix}} can be easily migrated to different SVS systems.
    \item We evaluate the proposed strategy on different SVS models. Extensive objective and subjective experiments demonstrate the effectiveness of \textit{\textbf{PHONEix}}, which brings significant improvement.
\end{itemize}




The rest of this article is organized as follows. Section 2 presents the proposed AFP strategy and phoneme distribution predictor. Section 3 presents our experimental setup, results, and discussion. Finally, Section 4 concludes the paper.

%% file: sec-method.tex
\section{Methodology}
\label{sec: method}

This section introduces our proposed AFP strategy \textit{\textbf{PHONEix}} and its integration with the SVS framework. 

\subsection{Overall Framework}
\label{ssec: overall framework}

As illustrated in Fig.~\ref{fig:model}, the whole system accepts inputs of the music score at note level and learns phoneme duration information guided by annotations. There are five steps in the application of the acoustic model: 1) A \textit{Prior Encoder} is attached before the major encoder-decoder-based structure. It consumes music scores and generates hidden representations of the music score at the note level. 2) A proposed \textit{Phoneme Distribution Predictor} is integrated to predict the proportion of phonemes in each syllable-note pair according to the \textit{\textbf{Type 1}} score features. 3) Then, an \textit{Acoustic Encoder} converts the predicted duration with its corresponding phoneme and pitch into musical features at the phoneme level. 4) A \textit{Duration Predictor} forecasts the acoustic frame lengths for each phoneme and passes the value to the \textit{Length Regulator} for expansion. 5) Finally, the \textit{Decoder} transfers the frame-level features to the spectrum.

\subsection{Acoustic Feature Processing Strategy }
\label{ssec: Acoustic Feature Processing Strategy}
Fig.~\ref{fig:AFP} illustrates the difference between \textit{\textbf{PHONEix}} and other AFP strategies.
Some of the previous works calculate acoustic features based solely on time sequence from the music score without referencing actual singing (Fig.~\ref{fig:input}~(a)). Others input the actual phoneme time sequence for training, which varies from the inference \textit{\textbf{Type 1}} input.
To bridge the gap between the music score and the actual singing voice (illustrated in Fig.~\ref{fig:model}), we employ a new AFP strategy \textit{\textbf{PHONEix}}. The input of the \textit{Acoustic Encoder} consists of the \textit{\textbf{Type 1}} note-level score features, \textit{i.e.,} phoneme, pitch, and note duration. Then, the final output of the \textit{Acoustic Encoder} is expanded to fit the actual duration under the guidance of an annotated phoneme time sequence. Nevertheless, the duration distributions of vowels and consonants vary sensibly. It is difficult for the duration predictor to provide an accurate frame expansion length with the same duration input among all phonemes in each note. In order to obtain a more precise phoneme pronunciation duration, a \textit{Prior Encoder} is introduced to help learn the proportion of phoneme pronunciation. The pronunciation of lyrics varies with pitch and note duration and is influenced by the context of the song. Hence, we utilize a Bi-LSTM or Transformer-based encoder instead of just analyzing linguistic features and providing predictions syllable by syllable. The \textit{Prior Encoder} can share the same networks as the \textit{Acoustic Encoder}, which can be easily adapted to various encoder-decoder-based acoustic models.

\subsection{Phoneme Distribution Predictor}
 As discussed in Sec.~\ref{ssec: overall framework}, we propose a learned phoneme duration for the following parts of the model, instead of using the same note duration for vowels and consonants or fixed phoneme duration (i.e. phoneme annotation). The learned phoneme duration is predicted by a proposed \textit{Phoneme Distribution Predictor} to distinguish vowels and consonants. To be specific, the predictor employs a series of 1-D convolutional layers to extract context information from the musical score features generated by the \textit{Prior Encoder}. The predictor gives a $n$-dimension probability distribution $\textbf{\textit{p}} = (p_1, p_2,\dots,p_n)$, where $n$ is the maximum number of phonemes in a syllable-note pair, which is determined by a syllable-phoneme lexicon. After that, we utilize the distribution to split the note duration $d^{\text{note}}$ extracted from the music score. The proportion for the $i$\textsuperscript{th} phoneme is $p_i$ and the annotated phoneme duration $d_i^{\text{ph}}$ is conform to actual phoneme distribution. 
A mean square error serves as an additional loss term to fit the predicted proportion to the real proportion in the actual singing pronunciation:
\begin{equation}
\mathcal{L}_{\text{ph}}=\Vert d^{\text{note}} \times \textbf{\textit{p}} - \textbf{\textit{d}}^{\text{ph}} \Vert_2.
\end{equation}

In some fast-speed songs, the phoneme duration can be too short to be passed by the acoustic frame resolution. To ensure integrity of phonemes, we assign a minimum duration to these short phonemes. Therefore, in inference, the phoneme duration in acoustic frame-level is defined as follows:
\begin{equation}
    \widehat{d}_i^{\text{ph}}=\mathrm{rint}[\max(d^{\text{note}} \times p_i, 1)] \quad (i=1,\dots,k),
\end{equation}
where $\mathrm{rint}$ is a rounding function that maps the duration to the closest integer.

%% file: sec-exp.tex
\begin{table*}[!t]
\small
\caption{ \small Comparison of the proposed AFP strategy, \textbf{\textit{PHONEix}}, with baselines (i.e., \textbf{\textit{Type 1}} and \textbf{\textit{Type 2}}). It is evaluated with Bi-LSTM or Transformer based encoder-decoder structures on Ofuton and Opencpop. Evaluations include three objective metrics (\textbf{MCD}, \textbf{VUV\_E}, and \textbf{SA}) and a subjective metric (\textbf{MOS}) are described in Sec.~\ref{ssec: exp settings}. The details of models and datasets are discussed in Sec.~\ref{ssec: dataset}. }
\centering

\begin{tabular}{c|clcccc}
\toprule
\textbf{Dataset}          & \textbf{Model}               & \multicolumn{1}{c}{\textbf{Method}} & \textbf{MCD ↓}            & \textbf{VUV E ↓}          & \textbf{SA ↑} & \textbf{MOS ↑} \\ 
\midrule
\multirow{8}{*}{Ofuton}   & \multirow{3}{*}{LSTM}         & \textit{\textbf{Type 1}}                         & 6.74          & \textcolor{white}{0}2.49\%           & 56.65\%            & 2.54 ± 0.05    \\
                          &                              & \textit{\textbf{Type 2}}                    & 6.34          & \textcolor{white}{0}2.53\%                    & 57.58\%            & 2.58 ± 0.05    \\
                          &                              & \textbf{\textit{\textbf{PHONEix}}}                    & \textbf{6.28} & \textcolor{white}{0}\textbf{2.41\%}           & \textbf{61.17\%}   & \textbf{3.03} ± 0.06    \\
\cmidrule(l){2-7}
                          & \multirow{3}{*}{Transformer} & \textit{\textbf{Type 1}}                         & 6.95          & \textcolor{white}{0}1.93\%                    & 61.09\%            & 2.60 ± 0.05    \\
                          &                              & \textit{\textbf{Type 2}}                    & 6.78          & \textcolor{white}{0}\textbf{1.71\%}           & 58.44\%            & 2.26 ± 0.05    \\
                          &                              & \textbf{\textit{\textbf{PHONEix}}}                    & \textbf{6.52} & \textcolor{white}{0}2.15\%           & \textbf{62.47\%}   & \textbf{3.01} ± 0.06   \\
\cmidrule(l){2-7}
                          &                              & Ground Truth                                  & -                        & -               & -                                          & 4.47 ± 0.04   \\
\midrule
\multirow{8}{*}{Opencpop} & \multirow{3}{*}{LSTM}         & \textit{\textbf{Type 1}}                         & 9.33          & \textcolor{white}{0}6.79\%          & 46.92\%            & 2.09 ± 0.05    \\
                          &                              & \textit{\textbf{Type 2}}                    & 8.62          &                     11.32\%                   & 53.93\%            & 2.48 ± 0.05   \\
                          &                              & \textbf{\textit{\textbf{PHONEix}}}                    & \textbf{7.90} & \textcolor{white}{0}\textbf{6.05\%}         & \textbf{60.97\%}   & \textbf{3.56} ± 0.06  \\
\cmidrule(l){2-7}
                          & \multirow{3}{*}{Transformer} & \textit{\textbf{Type 1}}                         & 8.77          & \textcolor{white}{0}6.19\%                   & 52.16\%            & 2.52 ± 0.05   \\
                          &                              & \textit{\textbf{Type 2}}                    & 9.05          &                     10.59\%                 & 52.90\%            & 1.89 ± 0.04   \\
                          &                              & \textbf{\textit{\textbf{PHONEix}}}                    & \textbf{8.42} & \textcolor{white}{0}\textbf{5.97\%}  & \textbf{60.47\%}   & \textbf{3.03} ± 0.05  \\
\cmidrule(l){2-7}
                          &                              & Ground Truth                                  & -                        & -               & -                                        & 4.72 ± 0.03   \\ 
\bottomrule
\end{tabular}
\label{tab: main result}
\end{table*}
\section{Experiments}
\label{sec: exp}

\subsection{Dataset}
\label{ssec: dataset}
We conduct our experiments on two datasets in two different languages. Every song is sampled at 24 kHz. We extract 80-dimensional Mel spectrograms with a window length of 1200 samples and a shifting size of 300 samples. The spectrograms normalized with global mean and variance from the train set are used as the ground truth for acoustic model outputs. 

\noindent\textbf{\textit{Ofuton:}}
an open male singing corpus, consisting of 56 Japanese songs \cite{Ofuton}. We divide the dataset by songs, resulting in a training set (51), a valid set (5), and a testing set (5), respectively.\footnote{Since there is no official split of the data, we instead use the same split as in the previous work \cite{shi2022muskits}.} The annotation information of the dataset includes manually labeled phoneme duration and music scores in MusicXML format. The segments of songs are separated by the \texttt{<sil>}, \texttt{<pau>}, and \texttt{<br>} silence marks, which align with segmentations of rest notes and breath marks in MusicXML.

\noindent\textbf{\textit{Opencpop: }}
a public Chinese female singing corpus of 100 songs in total \cite{Opencpop}. The annotation of phoneme duration and music score are organized in Textgrids. We follow the segmentations of the official release.

\subsection{Experimental Settings}
\label{ssec: exp settings}
The experiments are conducted using the music processing toolkit Muskits \cite{shi2022muskits}, which is adapted from ESPnet \cite{watanabe2018espnet}.
More details about the codes can be found on github \footnote{https://github.com/A-Quarter-Mile/PHONEix}.

\noindent\textbf{\textit{Model architectures:}}
We verify the proposed AFP strategy on two acoustic models. Both models adopt the encoder-decoder structure and the duration predictor proposed in FastSpeech~\cite{FastSpeech}. The models also share the configuration of the embedding layers. To be specific, we utilize 384-dimensional embedding layers for lyrics, notes, and note durations. The encoder and decoder of the first model are both three-layer 256-dimensional bidirectional Long-Short Term Memory units (Bi-LSTM) following~\cite{baseline1}. The second network utilizes a Transformer structure from~\cite{lu2020xiaoicesing}. Apart from the additional loss for phoneme distribution predictor, we apply the same loss functions as ~\cite{baseline1} and ~\cite{lu2020xiaoicesing}. We use the same HiFi-GAN vocoder~\cite{kong2020hifi} to convert the predicted acoustic features into waveforms.

\noindent\textbf{\textit{Training:}}
For both models, we utilize the Adam optimizer with a learning rate of 0.001 without schedulers. The batch size is 16. In all experiments,  we select the model checkpoints with the best validation losses for further evaluation.

\noindent\textbf{\textit{Evaluation metric:}}
We follow the objective and subjective evaluations in~\cite{baseline1}. For objective evaluation, we utilize Mel-cepstral distortion (\textbf{MCD}), voiced/unvoiced error rate (\textbf{VUV\_E}), and semitone accuracy (\textbf{SA}). For subjective evaluation, we randomly select 30 singing segments from the test set and invite 40 judges to score the quality of each segment on a scale of 1 to 5. We report the mean opinion scores (\textbf{MOS}) with a 95\% confidence interval.

\begin{table}[t]
\small
\caption{\small Objective evaluations of the ablation study on learned phoneme duration. Detailed discussions can be found in Sec.~\ref{ssec: ablation study}.}
\centering
\begin{tabular}{lcccc}
\toprule
\multicolumn{1}{c}{\textbf{Duration Source}}          & \textbf{MCD ↓}            & \textbf{VUV\_E ↓}         & \textbf{SA ↑}        \\
\midrule
\textit{\textbf{\textit{\textbf{PHONEix}}}} & \textbf{6.28 } & \textbf{2.41\%}          & 61.17\%          \\
\midrule
Annotation                      & 6.46          & 2.51\%                & \textbf{61.23\%} \\
Statistical Rule                   & 6.39           & 2.59\%          & 60.49\%          \\
Music Score                               & 6.73          & 2.69\%               & 59.03\%          \\
\midrule
Music Score (Note)                             & 6.79           & 2.74\%               & 58.39\%          \\
\bottomrule
\end{tabular}
\label{tab: learned phone duration}
\end{table}

                           

\subsection{Comparison with baselines}

We compare the proposed strategy with the \textit{\textbf{Type 1}} and \textit{\textbf{Type 2}} processing strategies described in Sec.~\ref{sec:intro}.
Table~\ref{tab: main result} presents the evaluation scores of the proposed acoustic processing strategy and the two baselines. All strategies are conducted with LSTM and Transformer-based encoder-decoder acoustic model structures on the Ofuton and Opencpop datasets, respectively.

\noindent
\textit{\textbf{Type 1}}: The AFP strategy of music score does not involve actual phoneme time sequence in the acoustic model. We apply the processing pipelines in Xiaoice~\cite{lu2020xiaoicesing}. It accepts score features at note level and expands frame lengths according to the phoneme duration produced by HMM-based forced-alignment~\cite{mcauliffe2017montreal,HMM}.

\noindent
\textit{\textbf{Type 2}}: The other way of processing acoustic features is to use the annotated phoneme time sequence as acoustic encoder input and length expansion ground truth for the length regulator. We analyze the acoustic feature following~\cite{baseline1}. The training stage has been tuned to fit the actual singing. However, the music score input in inference differs from the 
annotated ones in training as described in Sec. \ref{sec:intro}.

The results in Table~\ref{tab: main result} show significant improvements in both subjective and objective metrics in our SVS system equipped with an AFP strategy. The cases of corresponding results are shown in Fig.~\ref{fig:case}. The result of Fig.~\ref{fig:case} shows that the \textit{\textbf{PHONEix}} obtains the best estimation of duration of phonemes among all method, which further confirms the effectiveness of our proposed \textit{\textbf{PHONEix}}.

\subsection{Ablation study}
\label{ssec: ablation study}

In this subsection, we test the effectiveness of learnable phoneme duration input for the acoustic encoder and the validity of the proposed phoneme distribution predictor.

\begin{figure}[t]
\
	\centering
	\includegraphics[width=1.0\columnwidth]{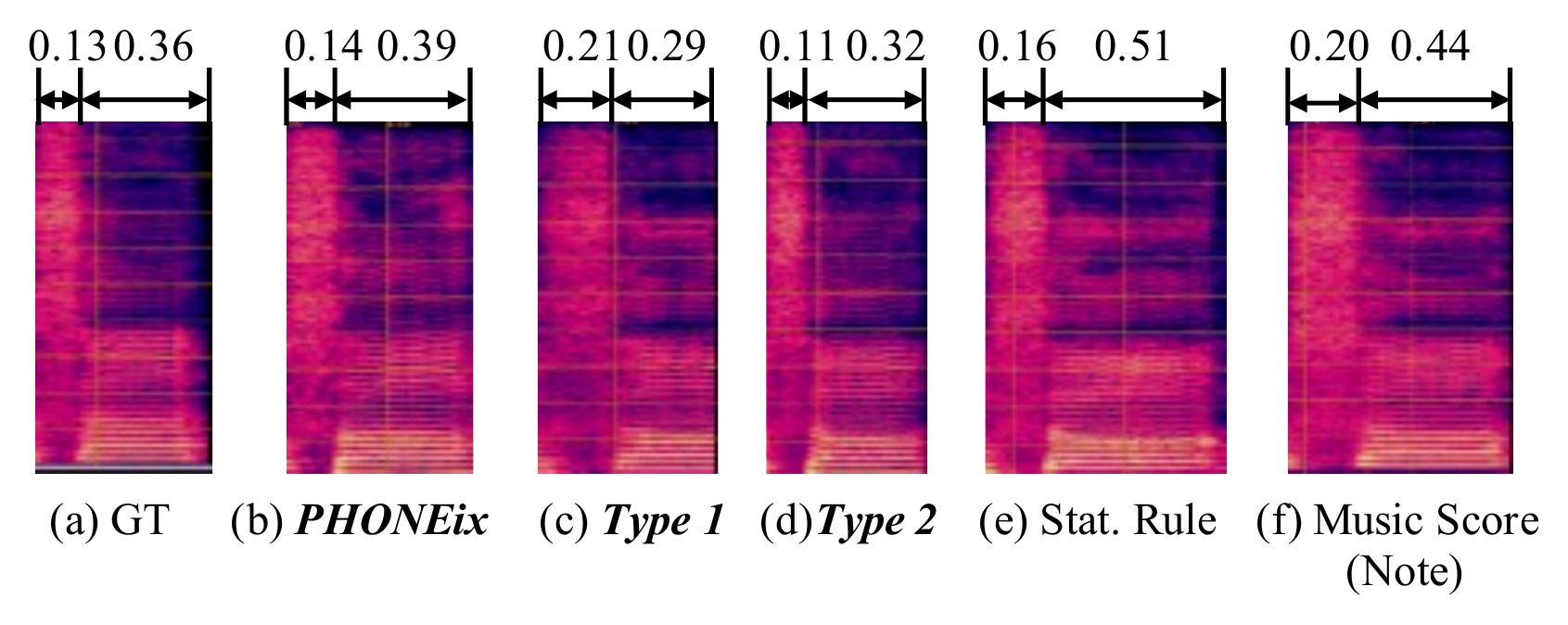}
 \vspace{-20pt}
	\caption{\small The cases of various methods. The number above the picture indicates the duration of the phoneme (in seconds). }

	\label{fig:case}
 \vspace{-10pt}
\end{figure}

In order to prove the effectiveness of learned phoneme duration, we compare it with other fixed duration sources for the acoustic encoder: 1) annotation (i.e., annotated phoneme duration). 2) statistical rule where the duration is derived from the statistical calculation of the corresponding dataset. 3) music score where the duration is obtained from the note duration. We use the same ground truth phoneme duration (annotated phoneme duration) in the \textit{Length Regulator} for the above three settings. The results show that the learned phoneme duration from \textit{\textbf{PHONEix}} gets a higher score in MCD and VUV\_E and comparable semitone accuracy to the annual annotation. The learned phoneme duration indicates the precise division of vowels and consonants in each note, which leads to ample pronunciation in singing. In addition, we train the SVS system that direct optimized over acoustic features at the syllable-note level (i.e. Music Score (Note) in Table~\ref{tab: learned phone duration}) and expand frame length by note time sequence. The note-level feature gets worse performances that it cannot generate the representation needed for good pronunciation without phoneme division (see Fig.~\ref{fig:case} (f)).


